\documentclass[letterpaper]{article} % DO NOT CHANGE THIS
\usepackage{aaai24}  % DO NOT CHANGE THIS
\usepackage{array}
\usepackage{latexsym}
\usepackage{graphicx}
\usepackage{subfigure}
\usepackage{enumitem}
\setlist[itemize]{itemsep=0pt, parsep=0pt, leftmargin = 0pt}
\usepackage{amsthm,amsmath,amssymb}
\usepackage{mathrsfs}
\usepackage{xcolor}
\usepackage{times}  % DO NOT CHANGE THIS
\usepackage{helvet}  % DO NOT CHANGE THIS
\usepackage{courier}  % DO NOT CHANGE THIS
\usepackage[hyphens]{url}  % DO NOT CHANGE THIS
\usepackage{graphicx} % DO NOT CHANGE THIS
\usepackage{natbib}  % DO NOT CHANGE THIS AND DO NOT ADD ANY OPTIONS TO IT
\usepackage{caption} % DO NOT CHANGE THIS AND DO NOT ADD ANY OPTIONS TO IT
\usepackage{algorithm}
\usepackage{algorithmic}
\usepackage{newfloat}
\usepackage{listings}
\DeclareCaptionStyle{ruled}{labelfont=normalfont,labelsep=colon,strut=off} % DO NOT CHANGE THIS
\floatstyle{ruled}
\newfloat{listing}{tb}{lst}{}
\floatname{listing}{Listing}
\title{FacetCRS: Multi-Faceted Preference Learning for Pricking Filter Bubbles in Conversational Recommender System}
\author{
    {Yongsen Zheng}\textsuperscript{\textnormal{1}},
    {Ziliang Chen}\textsuperscript{\textnormal{2}},
    {Jinghui Qin}\textsuperscript{\textnormal{3}},
    {Liang Lin}\textsuperscript{\textnormal{1}}\thanks{Liang Lin is the corresponding author.}\
}

\affiliations{
    \textsuperscript{\rm 1}Sun Yat-sen University\\
    \textsuperscript{\rm 2}Jinan University\\
    \textsuperscript{\rm 3}Guangdong University of Technology\\
     \{z.yongsensmile, scape1989\}@gmail.com,
     c.ziliang@yahoo.com, 
     linliang@ieee.org
}

\usepackage{bibentry}
\begin{document}

\maketitle

\begin{abstract}
The filter bubble is a notorious issue in Recommender Systems (RSs), which describes the phenomenon whereby users are exposed to a limited and narrow range of information or content that reinforces their existing dominant preferences and beliefs. This results in a lack of exposure to diverse and varied content. Many existing works have predominantly examined filter bubbles in static or relatively-static recommendation settings. However, filter bubbles will be continuously intensified over time due to the feedback loop between the user and the system in the real-world online recommendation. To address these issues, we propose a novel paradigm, Multi-Facet Preference Learning for Pricking Filter Bubbles in Conversational Recommender System (FacetCRS), which aims to burst filter bubbles in the conversational recommender system (CRS) through timely user-item interactions via natural language conversations. By considering diverse user preferences and intentions, FacetCRS automatically model user preference into multi-facets, including entity-, word-, context-, and review-facet, to capture diverse and dynamic user preferences to prick filter bubbles in the CRS. It is an end-to-end CRS framework to adaptively learn representations of various levels of preference facet and diverse types of external knowledge. Extensive experiments on two publicly available CRS-based datasets demonstrate that our proposed method achieves state-of-the-art performance in mitigating filter bubbles and enhancing recommendation quality in CRS.
\end{abstract}

\section{Introduction}
Unlike the traditional static recommendations \cite{TRS_1,TRS_2,TRS_new,IRS_1, IRS_2}, Conversational Recommendation Systems (CRSs) have emerged as powerful tools for providing personalized recommendations through natural language conversations and dialogue interactions \cite{dialogue_1,dialogue_2,dialogue_3}, which are broadly adopted in various domains, including online e-commerce \cite{e_commerce}, music recommendation \cite{music_conversation}, health counseling \cite{health}, \emph{etc}. However, CRSs frequently encounter the challenge of filter bubbles, wherein users are consistently presented with a restricted range of information and suggestions that are aligned with their dominant preferences throughout their interactions with the system \cite{CIRS}. Therefore, it is crucial to prick filter bubbles for improving performance of the CRSs.\\
\indent Recently, many research efforts have been dedicated to exploring filter bubbles in offline recommendation without considering user-item interactions \cite{diverse_user_preference_1,diverse_user_preference_2,diverse_user_preference_3,diverse_user_preference_4,diverse_user_preference_5,diverse_user_preference_7}.  These offline recommendation approaches primarily focus on investigating the underlying causes of filter bubbles. Through extensive experiments conducted on large-scale recommender systems, two primary factors have been discovered. The first primary factor is that users with less diverse preferences are more susceptible to becoming entrapped within filter bubbles. Another primary factor is that the learning mechanisms employed by the recommender amplify the filter bubble phenomenon as they tend to prioritize and accentuate a user's dominant interests.

Although these studies provide valuable insights into the phenomenon of filter bubbles, they have not adequately addressed the negative impacts of the feedback loop between the user and the system on the exacerbation of filter bubbles. More recently, Gao et al. \cite{CIRS} have made attempts to mitigate filter bubbles in Interactive Recommender Systems (IRSs) by incorporating user-system interactions over time through offline reinforcement learning, but this strategy cannot allow users to interact with the system through natural language to express their true thoughts.\\
\indent Despite their effectiveness, most existing methods still suffer from two major limitations: 1) \emph{Interactive Strategy}. Most methods primarily focus on addressing filter bubbles in offline recommendations in the static settings while overlooking the effect of the user-system feedback loop on filter bubbles. In reality, filter bubbles will be exacerbated when the user chats with the system over time \cite{filter_bubble,CIRS}. Even though a recent study CIRS \cite{CIRS} has emerged to combat filter bubbles interactively, its interaction strategy remains limited largely to rigid and single forms (\emph{e.g.}, click or skip, like or dislike). This fails to allow users to express their diverse, dynamic, and complicated preferences via natural language utterances, making it not applicable to real-world conversational recommendations. 2) \emph{Preference Exploration}. Most previous works \cite{diverse_user_preference_1,diverse_user_preference_2,diverse_user_preference_3,diverse_user_preference_4} have found that modeling users' diverse preferences can effectively prick filter bubbles. Based on this foundation, many methods strive to learn multi-level preferences to alleviate the filter bubble problem. But these approaches heavily rely on a limited set of sparse data sources such as historical user-item interactions, interest tags, and implicit and explicit ratings, in which a large portion of the available entries are empty or missing. This reliance on homogeneous and sparse data poses a significant limitation in excavating diverse and personalized preferences as it is still restricted by past preferences. Thus, it is hard to prevent filter bubbles from these scarce data sources.
\\
\indent To address these problems, we propose a novel framework, Multi-\textbf{Facet}ed Preference Learning for Pricking Filter Bubbles in \textbf{C}onversational \textbf{R}ecommender \textbf{S}ystem (\textbf{FacetCRS}). It models diverse user preferences from multiple facets adaptively, including entity-, word-, context-, and review-facet, to mitigate filter bubbles in the CRS. In these facets, the entity facet is extracted from item-oriented Knowledge Graph (KG) DBpedia \cite{DBpedia}. It contains all item entities appearing in the conversation based on the paths in such KG. This facet not only models user-item relationships better but also allows the recommender to discover new correlated items by the connections over KG, thus enabling the provision of diverse recommendations. Meanwhile, analyzing the prominent words employed by users during their interactions allows for identifying niche or specific preferences that might not be readily apparent through other facets of user profiling. Thus, we incorporate a word-oriented KG ConceptNet \cite{ConceptNet} to model the relations such as synonyms and co-occurrence between words. Moreover, the ongoing dialogue (\emph{i.e.}, context-facet preference) provides an opportunity to understand the evolving interests of users, so the CRSs can adapt their recommendations to align with current user preferences. Lastly, the reviews facet provides a direct and authentic information source reflecting the firsthand experiences of users. They often contain descriptive narratives, ratings, and opinions that shed light on various aspects of the items, such as quality, functionality, and performance. Analyzing these reviews allows recommender systems to identify common patterns, sentiment trends, and user sentiments towards specific features, helping them better understand user interests. We conduct experiments on two public CRS-based benchmarks, where our superior performance demonstrates the appealing reliability of our method.

Overall, our main contributions are included as follows:
\begin{itemize}
\item To the best of our knowledge, this is the first work to model user preference into multi-facets in the CRS, including entity-, word-, context-, and review-facet, to capture diverse user preferences to prick filter bubbles in the CRS.
\item We proposed a novel paradigm, FacetCRS, which accommodates multi-faceted user preference into an end-to-end CRS framework. It will adaptively learn representations of various levels of preference facet and multi-aspect knowledge.
\item Quantitative and qualitative experimental results on two CRS-based datasets exhibits superior performance of our proposed method, which serve as compelling evidence of the effectiveness of our FacetCRS in mitigating the filter bubble in the CRS.
\end{itemize}

\section{Related Work}
\subsection{Conversational Recommender System}
With the widespread adoption of intelligent agents in diverse domains \cite{related_work_dialogues}, conversational recommender systems \cite{ CRS_2,CRS_3,CRS_5} have attracted a lot of attention. These CRSs can be discerned into two primary categories: attribute-based CRS and human-like CRS. The former \cite{CRS_1,TCRS,CRS,LHID,multi-choice} focus on eliciting user preferences by prompting them to express their likes or dislikes with respect to various attributes, leveraging predefined actions such as item attributes and intent slots. Conversely, the latter \cite{RevCore,C2CRS,CRS_8,Topic-Guided,CRS_6,INSPIRED,TDCR,CRS_7} endeavors to furnish more lifelike recommendations through human-like responses. Human-like CRS typically comprises a conversation module responsible for generating appropriate responses and a recommendation module for delivering recommendations. Despite their sophistication, these approaches still suffer from the scarcity and insufficiency of contextual information present in initial conversational utterances. To address these challenges, existing methods often incorporate additional information from external sources, such as structured data like knowledge graphs \cite{C2CRS,zhou2020improving}, or unstructured data like reviews \cite{RevCore}, to enhance the conversation utterances. Nevertheless, they still grapple with drawbacks associated with a limited scope of knowledge facets in external sources. Our work adopts a human-like CRS approach and models user preferences across multiple facets, encompassing entities, words, contexts, and reviews.

\subsection{Filter Bubbles in Recommendation}
The filter bubble has emerged as a widely recognized predicament in the field of recommendation systems. When users find themselves ensnared within a filter bubble, the system tends to present information that aligns closely with their existing beliefs or interests, resulting in a curtailed exposure to a broader range of information \cite{diverse_user_preference_2,diverse_user_preference_3}. Recent research endeavors have delved into investigating the primary factors contributing to the formation of filter bubbles. From a user-centric perspective, individuals with narrower and less diverse preferences are more susceptible to becoming trapped within these bubbles \cite{diverse_user_preference_4,diverse_user_preference_5}. From a system perspective, the learning process can inadvertently amplify a user's dominant interests. Furthermore, the system typically assumes that user satisfaction equates to intrinsic interest by implicitly assuming that excessive exposure to items of interest will not affect user satisfaction \cite{CIRS,breakingfilterbubble}. For example, Wang et al. \cite{usercontrollable} introduce a novel recommender paradigm that empowers users to actively control the mitigation of filter bubbles. However, existing strategies predominantly concentrate on the static recommendation setting, which poses challenges in effectively modeling the dynamic nature of filter bubbles. In contrast, our approach models multifaceted user preferences when the user interacts with the system via natural language conversations to prick filter bubbles. 

\section{FacetCRS}
\begin{figure*}[ht]
    \centering
    \includegraphics[width=1\textwidth]{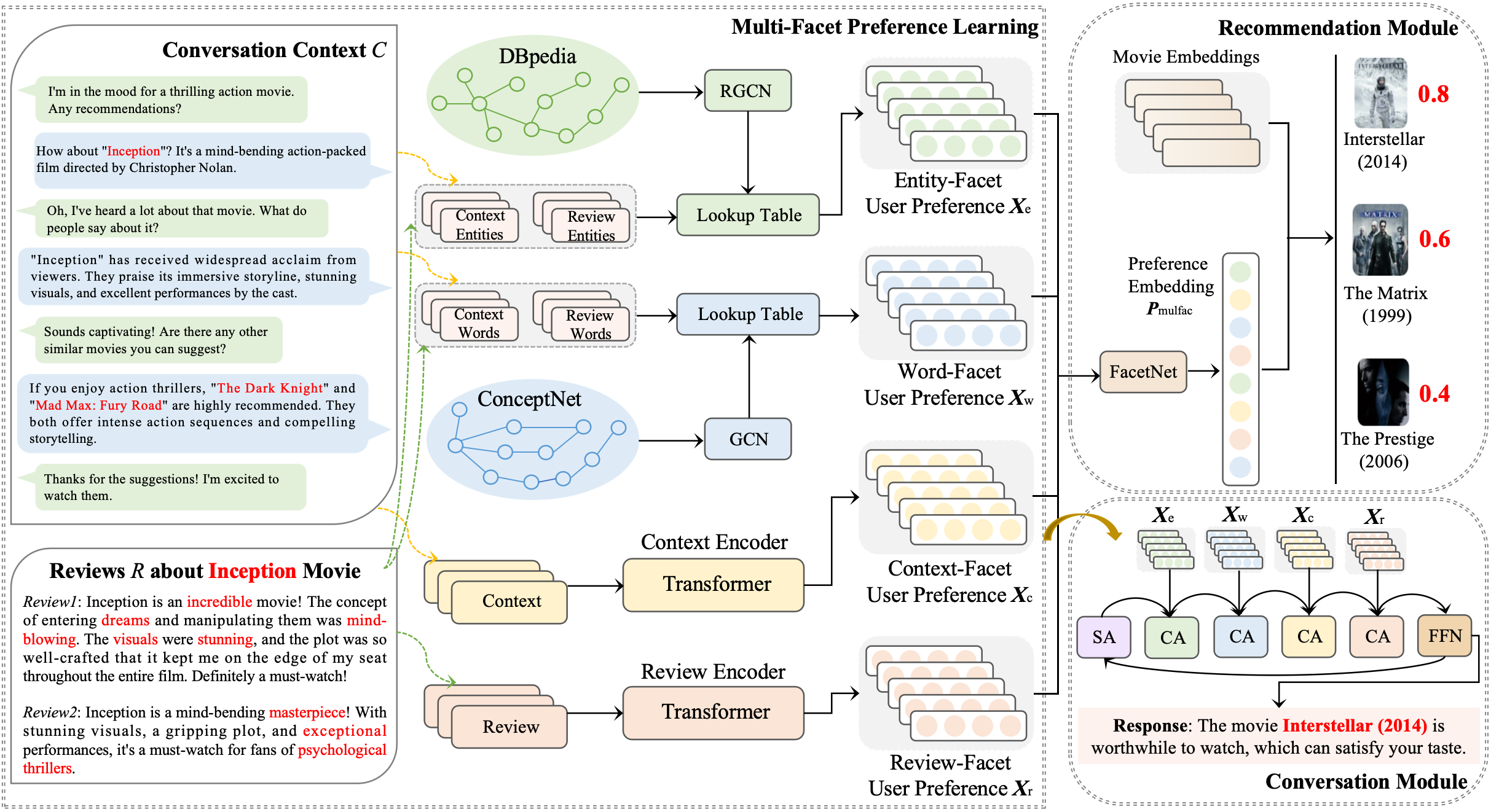}
    \caption{Overview of the proposed framework, FacetCRS, which consists of Multi-Facet Preference Learning and Multi-Facet Conversational Recommender System. The former aims to adaptively model diverse user preferences into multi-facets, including entity- , word-, context-, and review-facet, while the latter contains the recommendation module to select accurate items for the user and the conversation module to generate proper responses.}
    \label{fig:framework}
\end{figure*}

Filter bubble is a challenging issue in CRSs, and it will be increasingly severe over time due to the feedback loop between the user and the system in the online recommendation. To address these issues, we propose a novel framework, FacetCRS, which comprises Multi-Facet Preference Learning and Multi-Facet Conversational Recommender System. The pipeline of our FacetCRS is shown in Fig. \ref{fig:framework}. 

\subsection{Multi-Facet Preference Learning}
Numerous extensive experiments conducted by most existing recommendation methods \cite{diverse_user_preference_1,diverse_user_preference_2,diverse_user_preference_3} have consistently demonstrated that individuals with narrower preferences tend to be more susceptible to the confinement of filter bubbles. Along this line, we devise the Multi-Facet Preference Learning to model diverse user preferences into multi facets, including entity-, word-, context-, and review-facet, to improve recommendation diversity.
\subsubsection{Entity-Facet User Preference.} 
The KG entities facet not only enhance the user-item relationship modeling but also allows the recommender to discover new items by leveraging the connections within the KG, thus enabling the provision of diverse recommendations. To do this, we adopt the large-scale item-oriented KG DBpedia \cite{DBpedia} to extract the entities of the paths in the KG, in which those entities usually appear in conversations as items, inspired by \cite{RevCore,usercentric,C2CRS}. The entities always present in the form of head entities $e_{\rm head}$ and tail one $e_{\rm tail}$ in the fruitful knowledge triples $(e_{\rm head}, r, e_{\rm tail})$, where and $r$ is the relation between $e_{\rm head}$ and $e_{\rm tail}$. Given a conversation context $\mathcal{C} = \{s_t\}^n_{t=1}$, we first extract $k$ entities $\mathcal{E}^{c}_{k}=\{e_1, e_2, \cdots, e_k\}$ via DBpedia, where $s_{t}=\{w_j\}^m_{j=1}$ denotes one sentence (\emph{i.e.}, each utterance) and $\emph{w}$ represents a word from vocabulary $\mathcal{V}$. Then, we utilize the Relational Graph Convolutional Networks (RGCN) to learn entity embeddings. Concretely, we utilize RGCN to learn entity representations on the extracted subgraph, and thus the representation of entity $e$ at $(l+1)^{th}$ layer is calculated as:
\begin{equation}
\boldsymbol{e}^{l+1} = \sigma (\sum_{r \in \mathcal{R}} \sum_{\hat{e} \in \mathcal{H}^r_{e}} \frac{1}{\textsf{z}_{l+1}} \boldsymbol{W}^l_r {\hat{\boldsymbol{e}}}^{l} + \boldsymbol{W}^l \boldsymbol{e}^{l}).
\end{equation}Here ${\boldsymbol{e}}^l$ means the $l^{th}$ layer's representation of entity $e$, $\sigma$ is the sigmoid function, $\hat{e}$ is the entities from one-hop neighbor set $\mathcal{H}^r_{e}$ of entity $e$ under the relation $r$, ${\textbf{z}_{l+1}}$ is the hyperparameter, \emph{i.e.}, the normalization factor. $\boldsymbol{W}^l_r$ and $\boldsymbol{W}^l$ can be trained during model training. With the spirit of previous works \cite{usercentric,RevCore}, the last layer's representation $\boldsymbol{e}^L$ is employed as the entity representation $\boldsymbol{e}$ of the entity $e$. Therefore, the entity-facet embeddings $\boldsymbol{X}_{\rm e}$ can be described as:
\begin{equation}
\boldsymbol{X}_{\rm e} = {\rm RGCN}(\mathcal{E}^{c}_{k}) = \{\boldsymbol{e}^T_1, \boldsymbol{e}^T_2, \cdots, \boldsymbol{e}^T_k\}, 
\end{equation}where $\boldsymbol{e}^T_i$ is the embedding of the entity $e_i$ via RGCN. 
\subsubsection{Word-Facet User Preference.}
Keywords used in conversations serve as critical indicators to gain valuable insights into the user's specific needs. Analyzing the prominent words employed by users during their interactions allows for the identification of niche or specific preferences that might not be readily apparent through other facets of user profiling. Thus, word-facet information enables the delivery of more personalized and tailored recommendations. To model user preference in word facet, the external lexical word-oriented KG ConceptNet \cite{ConceptNet} is utilized to establish semantic relationships among words, encompassing diverse associations such as synonymy, antonyms, and co-occurrence. Given a context $\mathcal{C}$, the first step is to draw $m$ words $\mathcal{W}^c_m = \{w_1, w_2, \cdots, w_m\}$, and then utilize Graph Convolutional Networks (GCN) to learn the representations of current words. In GCN, the $(l+1)^{th}$ layer's representation of word $w$ can be expressed as:
\begin{equation}
\begin{aligned}
\boldsymbol{w}^{l+1} &= \sigma (\hat{\mathcal{A}} \boldsymbol{w}^{l} \boldsymbol{W}^{l}).
\end{aligned}
\end{equation}
where $\boldsymbol{w}^{l}$ is the $l^{th}$ layer's representation of word $w$, $\hat{\mathcal{A}}=\mathcal{D}^{-\frac{1}{2}}A\mathcal{D}^{-\frac{1}{2}}$ is the normalized adjacency matrix, $\mathcal{A}$ is the adjacency matrix by retrieving semantic relationships of words from ConceptNet, and $\mathcal{D}$ is a diagonal matrix of size $l_{\mathcal{A}} \times l_{\mathcal{A}}$ where the elements on the diagonal represent the degrees of the nodes, \emph{i.e.}, $\mathcal{D}_{ii} = \sum^{l_{\mathcal{A}}}_{j} \mathcal{A}_{ij}$. $\boldsymbol{W}^{l}$ is the trainable weights. Similar to RGCN, we use the last layer's representation $\boldsymbol{w}^L$ as the word representation $\boldsymbol{w}$ of the word $w$. Thus, the word-facet embeddings $\boldsymbol{X}_{\rm w}$ is given as:
\begin{equation}
\boldsymbol{X}_{\rm w} = {\rm GCN}(\mathcal{W}^{c}_{m}) = \{\boldsymbol{w}^T_1, \boldsymbol{w}^T_2, \cdots, \boldsymbol{w}^T_m\}, 
\end{equation}here $\boldsymbol{w}^T_i$ is the embedding of the word $w_i$ by GCN. 
\subsubsection{Context-Facet User Preference.}
Unlike static user profiles or explicit feedback, conversation contents (\emph{i.e.}, context-facet preference) provide a more nuanced view of user preferences by capturing the dynamic nature of their interests. The ongoing dialogue gives an opportunity to understand the evolving interests of users. By considering the conversational context, sentiment, and topics discussed, the CRS can adapt its recommendations to align with the user's current interests. Along this line, we employ a Transformer \cite{Trans} as the encoder to learn the representations of the conversation contexts. For the context $\mathcal{C}$, suppose the output embeddings of the previous transformer layer is $\mathcal{T}^{(l-1)}(\mathcal{C})$, then the current one $\mathcal{T}^{l}(\mathcal{C})$ can be defined by \emph{Multi-head Attention} function $\textsf{MHA}(\cdot,\cdot,\cdot)$ as follows:
\begin{equation}
\begin{aligned}
&\mathcal{T}^{l}(\mathcal{C})=\textsf{MHA}(\mathcal{T}^{l-1}(\mathcal{C}), \mathcal{T}^{l-1}(\mathcal{C}), \mathcal{T}^{l-1}(\mathcal{C})),\\
&\textsf{MHA}(\boldsymbol{K}, \boldsymbol{Q}, \boldsymbol{V}) = [\textsf{head}^{l}_{1};\cdots;\textsf{head}^{l}_{h}]\boldsymbol{W}^{l}_{j},
\label{multi_head_attention}
\end{aligned}
\end{equation}where $h$ is the number of heads, $\boldsymbol{W}^{l}_{j}$ is the training parameteres, and each head $ \textsf{head}^{l}_{j}$ is computed by 
\emph{Scaled Dot-Product Attention} \cite{allneed} functions $\textsf{SDA}(\cdot,\cdot,\cdot)$ as:
\begin{equation}
\begin{aligned}
 \textsf{head}^{l}_{j}=\textsf{SDA}(\mathcal{T}^{l-1}(\mathcal{C})\boldsymbol{W}^{k}_{j}, & \mathcal{T}^{l-1}(\mathcal{C})\boldsymbol{W}^{q}_{j}, \mathcal{T}^{l-1}(\mathcal{C})\boldsymbol{W}^{v}_{j}),\\
 \textsf{SDA}(\boldsymbol{K},\boldsymbol{Q},\boldsymbol{V})=&\textsf{Softmax}(\frac{\boldsymbol{Q}\boldsymbol{K}^{\rm T}}{\sqrt{d/h}})\boldsymbol{V}.
\end{aligned}
\label{SDPA}
\end{equation}
Note that $\forall j\in[h]$, $\boldsymbol{K}$,$\boldsymbol{Q}$ and $\boldsymbol{V}$ denote the key, query, and value matrices, respectively. $\textsf{Softmax}$ is the softmax function. $\boldsymbol{W}^{k}_{j}$, $\boldsymbol{W}^{q}_{j}$ and $\boldsymbol{W}^{v}_{j}$ are the parameters that can be learned. For convenience, we employ the output embedding of the last transformer layer to be context-facet embeddings $\boldsymbol{X}_{\rm c}$. Based on Eq.(\ref{multi_head_attention}), it can be written as:
\begin{equation}
\begin{aligned}
\boldsymbol{X}_{\rm c}=\textsf{MHA}(\mathcal{T}^{L-1}(\mathcal{C}), \mathcal{T}^{L-1}(\mathcal{C}), \mathcal{T}^{L-1}(\mathcal{C})).
\label{multi_head_attention_1}
\end{aligned}
\end{equation}
Here, $L$ denotes the number of transformer layers.
\subsubsection{Review-Facet User Preference.} 
Item reviews effectively reflect the firsthand experiences of users by describing the authentic user experiences and feelings. They often contain descriptive narratives, ratings, and opinions that shed light on various aspects of the items, such as quality, functionality, and performance. Analyzing these reviews is beneficial to identify common patterns, sentiment trends, and user sentiments towards specific features, helping them better understand user interests. Due to the advantages of reviews in exploring user preferences, some works \cite{RevCore,C2CRS} have introduced reviews to improve CRS. Inspired by them, we also incorporate reviews to explore user preferences in the review facet. Given a review $\mathcal{R}$ that is retrieved from a set of reviews based on \cite{RevCore}, we also adopt the Transformer to model the review representations to complement other three facets of modeling user preference. Similar to context-facet embeddings modeling, the output embeddings of the last transformer layer are considered as the review-facet user embeddings. It can be defined as follows according to Eq.(\ref{multi_head_attention}):
\begin{equation}
\begin{aligned}
\boldsymbol{X}_{\rm r}=\textsf{MHA}(\mathcal{T}^{L-1}(\mathcal{R}), \mathcal{T}^{L-1}(\mathcal{R}), \mathcal{T}^{L-1}(\mathcal{R})).
\label{multi_head_attention_2}
\end{aligned}
\end{equation}
\subsubsection{Fusing Multi-Facet User Preference.} 
In the preceding sections, we have formulated user preferences at various levels of detail, encompassing entity-, word-, context-, and review-facet considerations, gradually progressing from a broader perspective to a more nuanced one to model diverse user preference. These four types of preference facets differ in their definitions but are complementary to each other, allowing for diverse preference modeling and effectively mitigating filter bubbles in the CRS. We contend that user preferences may be collectively influenced by these facets, albeit with varying magnitudes. Consequently, the ultimate user preference representation $\boldsymbol{P}_{\rm mulfac}$ can be achieved by our proposed multifaceted fusion function FacetNet$(\cdot)$ as follows:
\begin{equation}
{\rm FacetNet}(\cdot)=
\left\{
\begin{aligned}
\mathcal{T}^{L}_{\rm enwo} &= \mathcal{T}^{L}(\boldsymbol{X}_{\rm e}) \otimes \mathcal{T}^{L}(\boldsymbol{X}_{\rm w}),\\
\mathcal{T}^{L}_{\rm core} &= \mathcal{T}^{L}(\boldsymbol{X}_{\rm c}) \otimes \mathcal{T}^{L}(\boldsymbol{X}_{\rm r}),\\
\mathcal{T}^{L}_{\rm mfac} &=\mathcal{T}^{L}_{\rm enwo} \oplus \mathcal{T}^{L}_{\rm core},\\
\boldsymbol{P}_{\rm mulfac}  &= \textsf{Softmax}(\textsf{ReLU}(\mathcal{T}^{L}_{\rm mfac})).\\
\end{aligned}
\right.
\end{equation}Here, the symbol $\otimes$ represents the matrix multiplication operation, while $\oplus$ signifies vector concatenation. $\textsf{ReLU}$ is the Rectified Linear Unit. Subsequently, the multifaceted user preference $\boldsymbol{P}_{\rm mulfac}$ is employed to provide accurate recommendations and generate suitable responses in the CRS.
\subsection{Multi-Facet Conversational Recommender System}
To prick filter bubbles in the CRS, we adopt the multifaceted user preference $\boldsymbol{P}_{\rm mulfac}$ to effectively make item predictions in the recommendation module and accurately predict the next utterances in the conversation module.
\subsubsection{Recommendation Module.}
The main goal of the recommendation module is to deliver precise item suggestions to users by leveraging their interactions with the system through natural conversations. In order to mitigate the impact of filter bubbles, we aim to explore the breadth of user preferences and encompass a diverse range of recommendations. Therefore, the multifaceted user preference $\boldsymbol{P}_{\rm mulfac}$ is used to excavate diverse user interests from multiple perspectives to enhance recommendation performance. Specifically, $\boldsymbol{P}_{\rm mulfac}$ is first fed into the Multilayer Perceptron Layer $(\textsf{MPLNet})$ followed by a softmax layer to generate the recommendation prediction $\textsf{P}_{\rm r}$. This process is:
\begin{equation}
\textsf{P}_{\rm r} = \textsf{Softmax}(\textsf{MPLNet}(\boldsymbol{P}_{\rm mulfac})).
\end{equation}
To train the model, we adopt the cross-entropy loss \cite{RevCore} to achieve our learning goals. The cross-entropy loss $\mathcal{L}_{\rm r}$ measures the dissimilarity between the prediction $P_{\rm r}$ and the target item category and can be computed as:
\begin{equation}
\mathcal{L}_{\rm r} = -\frac{1}{N_{\rm r}}\sum^{N_{\rm r}}_{j = 1} {\rm log} \textsf{P}^{j}_{\rm r},
\end{equation}
where $N_{\rm r}$ denotes total number of recommendations, and $P^{(j)}_{\rm r}$ is the target category in the $j$-th recommendation. 

\subsubsection{Conversation Module.}
The goal of the conversation task is to generate suitable utterances to respond to the user. In our conversation module, we adopt the entity-facet embeddings $\boldsymbol{X}_{\rm e}$, word-facet embeddings $\boldsymbol{X}_{\rm w}$, context-facet embeddings $\boldsymbol{X}_{\rm c}$, and review-facet embeddings $\boldsymbol{X}_{\rm r}$ to the decoder network for simultaneously predicting the next responses. Concretely, these multi-facet user preference embeddings are fed into each cross-attention layer to fuse their useful knowledge information to produce diverse response contents. Formally, we adopt $\boldsymbol{Y}^{(i-1)}$ to denote the output of the last time unit, then the current one $\boldsymbol{Y}^{(i)}$ is :
\begin{equation}
\begin{aligned}
\boldsymbol{A}^{i}_0 &= \textsf{MHA}(\boldsymbol{Y}^{i-1}, \boldsymbol{Y}^{i-1}, \boldsymbol{Y}^{i-1}),\\
\boldsymbol{A}^{i}_1 &= \textsf{MHA}(\boldsymbol{A}^{i}_0, \boldsymbol{X}_{\rm e}, \boldsymbol{X}_{\rm e}),\\
\boldsymbol{A}^{i}_2 &= \textsf{MHA}(\boldsymbol{A}^{i}_1, \boldsymbol{X}_{\rm w}, \boldsymbol{X}_{\rm w}),\\
\boldsymbol{A}^{i}_3 &= \textsf{MHA}(\boldsymbol{A}^{i}_2, \boldsymbol{X}_{\rm c}, \boldsymbol{X}_{\rm c}),\\
\boldsymbol{A}^{i}_4 &= \textsf{MHA}(\boldsymbol{A}^{i}_3, \boldsymbol{X}_{\rm r}, \boldsymbol{X}_{\rm r}),\\
\boldsymbol{Y}^{i} &= \textsf{FFN}(\boldsymbol{A}^{i}_4).
\label{decoder}
\end{aligned}
\end{equation}Here, $\textsf{MHA}$ can be computed by Eq. (\ref{multi_head_attention}), and the $\textsf{FFN}(\cdot)$ is the fully-connected feed-forward network, which can be modelled as follows:
\begin{equation}
\begin{aligned}
\textsf{FFN}(\boldsymbol{x}) = \textsf{ReLU}(\boldsymbol{x} \boldsymbol{W}_1+ \boldsymbol{b}_1)\boldsymbol{W}_2+ \boldsymbol{b}_2,
\label{FFN}
\end{aligned}
\end{equation}where $\boldsymbol{W}_1$ and $\boldsymbol{W}_2$ represent the learnable weight matrices, and $\boldsymbol{b}_1$ and $\boldsymbol{b}_2$ denote the bias terms corresponding to the first and second layers, respectively. As illustrated above, multifaceted user preference information is progressively integrated into the decoding stage. Initially, the entity-facet user preference is incorporated, then the word-facet user preference, followed by the context-facet user preference, and lastly, the review-facet user preference. Integrating multiple facets of information enriches users' understanding of items, resulting in a comprehensive and diverse range of user interests, which can effectively prick filter bubbles in CRS. To train the conversation module, we also adopt the cross-entropy loss inspired by \cite{RevCore} to learn the response generation. The conversation loss $\mathcal{L}_{\rm c}$ can be illustrated as:
\begin{equation}
\begin{aligned}
\mathcal{L}_{\rm c} &= - \frac{1}{M} \sum^M_{t=1} {\rm log}(\tilde{\textsf{P}}( s_t | \{s_{t-1}\})),\\
\tilde{\textsf{P}}( s_t | \{s_{t-1} \}) &= \tilde{\textsf{P}}_v(s_t | \textsf{Y}_i)+\tilde{\textsf{P}}_g(s_t | \textsf{Y}_i, \mathcal{G})+\tilde{\textsf{P}}_r(s_t | \textsf{Y}_i, \mathcal{R}),
\label{conversation_loss}
\end{aligned}
\end{equation}where $M$ is the number of conversation turns, $s_t$ represents the $t^{th}$ utterance in the conversation, $\{s_{t-1}\}$ represent the previously generated sub-sequence, which can be denoted as $\{s_{t-1}\}=s_1, s_2, \cdots, s_{t-1}$. $\tilde{\textsf{P}}( s_t | \{s_{t-1}\})$ refers to the generation probability of the next token $s_t$. We utilize $\tilde{\textsf{P}}_v(\cdot)$, $\tilde{\textsf{P}}_g(\cdot)$, and $\tilde{\textsf{P}}_r(\cdot)$ as probability functions over the vocabulary $\mathcal{V}$, entities from KG $\mathcal{G}$, and reviews $\mathcal{R}$ with $\textsf{Y}_i$ as the input. 

\section{Experiments and Analyses}
In this section, we conduct experiments to evaluate the performance of FacetCRS and answer the following questions:
\begin{itemize}
\item \textbf{RQ1:} How does FacetCRS perform compared with state-of-the-art methods in the recommendation task?
\item \textbf{RQ2:} How does FacetCRS perform compared with state-of-the-art methods in the conversation task?
\item \textbf{RQ3:} How does FacetCRS prick filter bubbles in the CRS?
\item \textbf{RQ4:} How do the entity-facet user preference $\boldsymbol{X}_{\rm e}$, word-facet $\boldsymbol{X}_{\rm w}$, context-facet $\boldsymbol{X}_{\rm c}$, and review-facet $\boldsymbol{X}_{\rm r}$ contribute to the performance?
\end{itemize}
\subsection{Experimental protocol}
\textbf{Datasets.} We evaluate our FacetCRS on two challenging CRS-based datasets \textbf{REDIAL}  \cite{TDCR} and \textbf{TG-REDIAL} \cite{Topic-Guided}. \textbf{REDIAL} encompasses a collection of 10,006 conversations associated with 51,699 movies. It is partitioned into three training, validation, and testing datasets, respectively. Additionally, this dataset incorporates a review database sourced from the IMDb\footnote{https://www.dbpedia.org/} website, containing 30 reviews for each movie. \textbf{TG-REDIAL} is a Chinese dataset consisting of 10,000 dialogues involving two parties, which collectively comprise a total of 129,392 utterances relevant to 33,834 movies. Each conversation commences with an initial sentence and unfolds chronologically, giving rise to subsequent response utterances and recommendations. The review database integrated within TG-REDIAL is obtained from the Douban\footnote{https://movie.douban.com/} website.\\
\noindent \textbf{Metrics.} To make a fair comparison, we use same metrics as most previous works \cite{RevCore,C2CRS}. For the recommendation task, we adopt Recall@\emph{k} (R@\emph{k}, with \emph{k} = 1, 10, 50). For the conversation task, we adopt both automatic evaluation (\emph{i.e.}, Distinct \emph{n}-gram (D-\emph{n}, with \emph{n} = 2, 3, 4)) and human evaluation (\emph{i.e.}, \emph{Fluency} and \emph{Informativeness}).\\
\noindent \textbf{Baselines.} To comprehensively evaluate the performance of our FacetCRS on two datasets, we conduct a comparative analysis with several state-of-the-art models in both recommendation and conversation tasks. The compared methods include \textbf{Popularity}, \textbf{TextCNN} \cite{TextCNN}, \textbf{Trans} \cite{Trans}, \textbf{ReDial} \cite{TDCR}, \textbf{KBRD} \cite{chen2019towards}, \textbf{KGSF} \cite{zhou2020improving}, \textbf{KECRS} \cite{KECRS}. By conducting a thorough comparison with these state-of-the-art models, we can effectively evaluate the performance and effectiveness of our FacetCRS in the recommendation and conversation tasks.

\subsection{Evaluation on Recommendation Task (RQ1)}
We utilize the Recall@\emph{k} to evaluate our FacetCRS in the recommendation task. Table \ref{tab:recommendation} summarizes the experimental results. We can see that our model is superior to all the compared methods. The best is labeled in boldface.\\
\indent The improvement of FacetCRS over these baselines can be attributed to three aspects: (1) In the real-world, the data of user interactions is rather sparse, and thus external data is crucial to complement the user behavior data to explore user preference. Thus, we adopt multi-aspect external data, such as DBpedia, ConceptNet, and item reviews, to model diverse user interests; (2) Besides these external data, the dialogue contexts between the user and the system in the CRS also play an important role in excavating users' dynamic interests and evolving needs. Hence, modeling multifaceted user preferences via this multi-dimension knowledge is beneficial to explore diverse user preferences, which can effectively make recommendations. (3) FacetCRS is designed to prick filter bubbles in the CRS when the user chats with the system over time via natural language conversations. To do this, we adopt the multifaceted user preference $\boldsymbol{P}_{\rm multfac}$ to select the important candidate items to recommend to the user in the recommendation module. In this manner, leveraging multifaceted user preferences can effectively prick filter bubbles in the CRS via the user-system feedback loop.

\subsection{Evaluation on Conversation Task (RQ2)}
\subsubsection{Automatic Evaluation.} Table \ref{tab:conversation} presents the D@\emph{n} results in our FacetCRS on the conversation task. The experimental results unequivocally demonstrate that our proposed FacetCRS attains superior performance compared to all competing methods, consistently outperforming them across various metrics. First, we can see that ReDial outperforms Popularity and TextCNN because it introduces a pre-trained RNN module. Meanwhile, KBRD is superior to KECRS since it incorporates external data (\emph{e.g.}, DBpedia) to align the item and word representations. Additionally, KGSF performs better than other baselines. This is attributed that KGSF not only aligns the conversation contexts and items but also further enhances their representations.\\
\indent Compared with these baselines, the improvement of FacetCRS can be attributed to several main reasons: (1) We adopt the entity-facet user preference $\boldsymbol{X}_{\rm e}$, word-facet $\boldsymbol{X}_{\rm w}$, context-facet $\boldsymbol{X}_{\rm c}$, and review-facet $\boldsymbol{X}_{\rm r}$ to model feature preferences at different latitudes, which is helpful of modeling user preference into multi-facet, thereby mitigating the filter bubbles in the CRS and make diverse recommendation. (2) To accurately predict the next utterances, we take these four facet user preferences into the cross-attention to obtain the respective context representations to complement each other for generating suitable responses. 

\begin{table}[tp]
     \small
	\setlength{\tabcolsep}{1.2mm}
	\setlength{\abovecaptionskip}{0pt}  
	\begin{center}
		\renewcommand{\arraystretch}{1.0}
  		\caption{Results on the recommendation task. Numbers marked with * indicate that the improvement is statistically significant compared with the best baseline ($t$-test with $p$-value $\textless$ 0.05).}
        \label{tab:each_component}
		\begin{tabular}{>{\raggedright\arraybackslash}p{1.5cm}|ccc|ccc}
             \hline
              {Datasets}&\multicolumn{3}{c|}{REDIAL} &\multicolumn{3}{c}{TG-REDIAL} \\
			\hline
		     Models&R@1&R@10&R@50 &R@1&R@10&R@50\\
			\hline
			Popularity & 0.011& 0.054&0.183&0.0004&0.003&0.014\\
			TextCNN& 0.013&0.068&0.191&0.003&0.010&0.024\\
              ReDial &0.024&0.140&0.320&0.000&0.002&0.013\\
              KBRD &0.031 &0.150&0.336&0.005 &0.032&0.077\\
              KGSF &0.039 &0.183&0.378&0.005 &0.030&0.074\\
              KECRS &0.021&0.143&0.340&0.002&0.026 &0.069\\
             \textbf{FacetCRS}&\textbf{0.041*} &\textbf{0.202*}&\textbf{0.386*}&\textbf{0.006*}&\textbf{0.034*}&\textbf{0.080*}\\
			\hline
		\end{tabular}
		\label{tab:recommendation}
	\end{center}
\end{table}

\begin{table}[tp]
     \small
	\setlength{\tabcolsep}{1.2mm}
	\setlength{\abovecaptionskip}{0pt}  
	\begin{center}
		\renewcommand{\arraystretch}{1.0}
  		\caption{Results on the conversation task via automatic evaluation. Numbers marked with * indicate that the improvement is statistically significant compared with the best baseline ($t$-test with $p$-value $\textless$ 0.05).}
        \label{tab:each_component}
		\begin{tabular}{>{\raggedright\arraybackslash}p{1.5cm}|ccc|ccc}
             \hline
              {Datasets}&\multicolumn{3}{c}{REDIAL} &\multicolumn{3}{c}{TG-REDIAL} \\
			\hline
		     Models&D-2&D-3&D-4 &D-2&D-3&D-4\\
			\hline
			Trans&0.067&0.139&0.227&0.053&0.121&0.204\\
              ReDial &0.082&0.143&0.245&0.055&0.123&0.215\\
              KBRD &0.086&0.153&0.265&0.045&0.096&0.233\\
              KGSF &0.114&0.204&0.282 &0.086&0.186&0.297\\
              KECRS &0.040&0.090&0.149&0.047&0.114&0.193\\
             \textbf{FacetCRS}&\textbf{0.126*}&\textbf{0.209*}&\textbf{0.305*}&\textbf{0.113*}&\textbf{0.228*}&\textbf{0.312*}\\
			\hline
		\end{tabular}
		\label{tab:conversation}
	\end{center}
\end{table}

\begin{table}[tp]
    \small
	\setlength{\tabcolsep}{6mm}
	\setlength{\abovecaptionskip}{0pt}  
	\begin{center}
		\renewcommand{\arraystretch}{1.0}
		\caption{Results on the conversation task via human evaluation. Numbers marked with * indicate that the improvement is statistically significant compared with the best baseline ($t$-test with $p$-value $\textless$ 0.05).}
        \label{tab:each_component}
		\begin{tabular}{>{\raggedright\arraybackslash}p{1.5cm}|cc}
             \hline
              Models&Fluency&Informativeness \\
			\hline
			Trans&0.97&0.92\\
              ReDial &1.35&1.04\\
              KBRD &1.23&1.15\\
              KGSF &1.48&1.37\\
              KECRS &1.39&1.19\\
             \textbf{FacetCRS}&\textbf{1.50*}&\textbf{1.39*}\\
			\hline
		\end{tabular}
		\label{tab:human_evaluation}
	\end{center}
\end{table}

\subsubsection{Human Evaluation.}
Table \ref{tab:human_evaluation} shows the experimental results in our FacetCRS via human evaluation on the conversation task. From the experimental results, we can observe that: 1) ReDial performs better than Transformer since it has a pre-trained RNN encoder; 2) KGSF outperforms multiple baselines such as ReDial, KBRD, KECRS in terms of \emph{informativeness} metric because it incorporates the external KG to align the word and item representations; 3) From the view of \emph{informativeness}, KGSF also performs better than all other baselines attributed to the fact that it enhances representations via KG.

FacetCRS consistently achieves superior performance in both metrics compared to all the baselines. Our model incorporates various facets, including entity facet, word facet, context facet, and review facet, thereby effectively mitigating filter bubbles in the CRS as the user interacts with the system over time. Additionally, our model generates conversation utterances that are both more informative and fluent via these four facets of knowledge information. 

\subsection{Study on Filter Bubbles (RQ3)}
Since we aim to prick filter bubbles in the CRS, we examine the recommendation results compared with the strongest baselines to verify whether FacetCRS can mitigate the filter bubbles. To this end, we adopt multiple metrics to fully evaluate the severity of filter bubbles.  

\emph{1) Isolation-Index (Iso-Index) \& Coverage}: In Table \ref{tab:filter_bubbles_}, we present the experimental results for two widely-used metrics, namely \emph{Iso-Index} and \emph{Coverage}. It is evident that our FacetCRS consistently demonstrates the lowest \emph{Iso-Index} values and the highest \emph{Coverage} values across both datasets. The reduced \emph{Iso-Index} values indicate a greater recommendation diversity, encompassing a broader range of facets and topics. Moreover, the higher \emph{Coverage} values achieved by FacetCRS confirm its ability to effectively cover a larger portion of the recommendation space. These observations highlight the capability of FacetCRS to effectively mitigate filter bubbles by enhancing the diversification of recommendations. 

\begin{figure}[h]
\centering
\subfigure[KECRS]{\label{fig:dimension_hr} \includegraphics[scale=0.2]{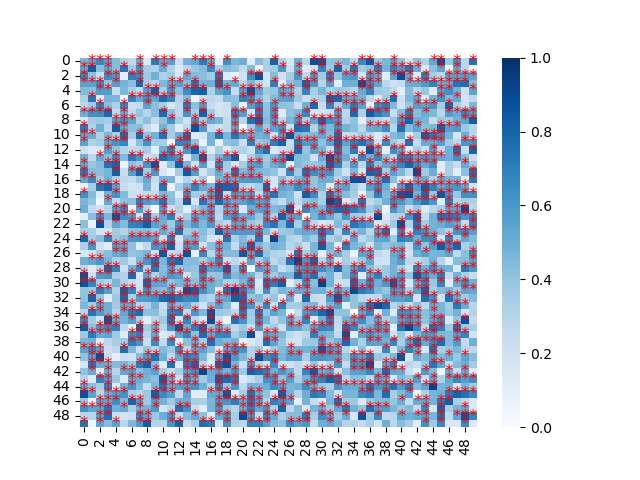}}
\subfigure[KBRD]{\label{fig:dimension_ndcg} \includegraphics[scale=0.2]{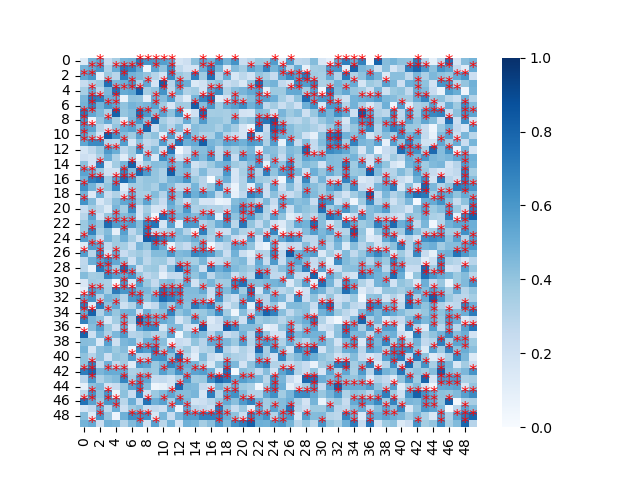}}
\subfigure[KGSF]{\label{fig:dimension_ndcg} \includegraphics[scale=0.2]{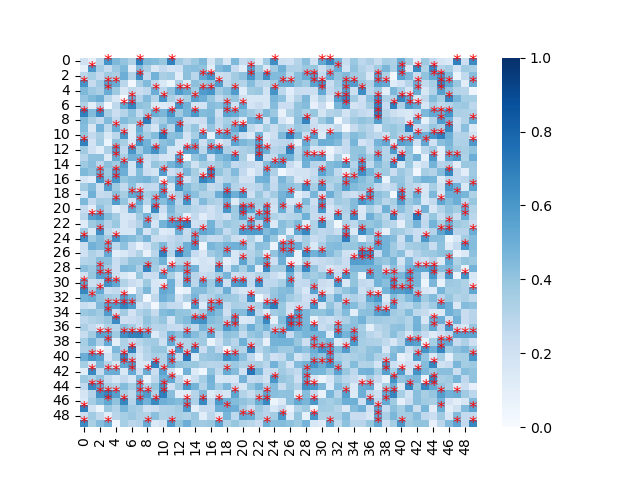}}
\subfigure[Ours]{\label{fig:dimension_ndcg} \includegraphics[scale=0.2]{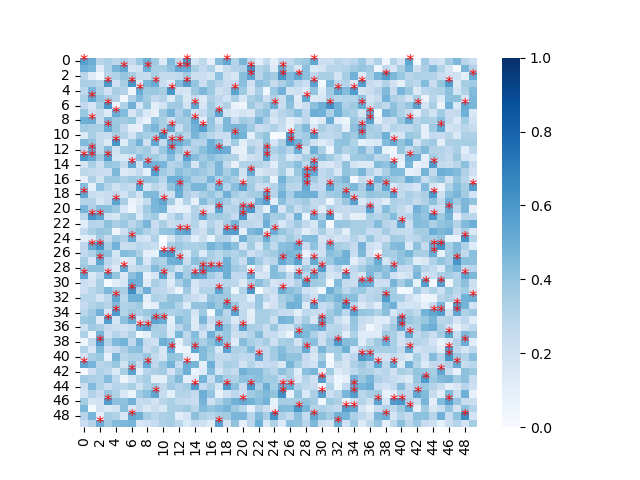}}
\caption{The heat maps depicting item similarity.}
\label{fig:heat_map}
\end{figure}

\emph{2) Item Similarity}: In Figure \ref{fig:heat_map}, we evaluate the similarity among the representations of the top 50 recommended items generated by each method on the REDIAL dataset. Heat map with brighter color serves as an indication of increased recommendation diversity while darker hues suggest a contrasting scenario. Additionally, the presence of more * markers indicates higher similarity and lower diversity. Analysis of the heat map reveals that KECRS exhibits a darker color and the highest * magnitudes, indicating lower diversity in its recommended items. Similar characteristics are observed in the KBRD and KGSF methods. Conversely, our approach, FacetCRS, demonstrates superior performance with a noticeably lighter heat map and fewer * markers. This serves as clear evidence of the effectiveness of FacetCRS in combatting filter bubbles by providing a wide range of diverse recommendations.

\begin{table}[t]
     \small
	\setlength{\tabcolsep}{1.2mm}
	\setlength{\abovecaptionskip}{0pt}  
	\begin{center}
		\renewcommand{\arraystretch}{1.0}
  		\caption{Results on \emph{Iso-Index} and \emph{Coverage} metrics.}
        \label{tab:each_component}
		\begin{tabular}{>{\raggedright\arraybackslash}p{1.5cm}|cc|cc}
             \hline
              {Datasets}&\multicolumn{2}{c|}{REDIAL} &\multicolumn{2}{c}{TG-REDIAL} \\
			\hline
		     Models&Iso-Index $\downarrow$&Coverage$\uparrow$&Iso-Index$\downarrow$&Coverage$\uparrow$\\
			\hline
			Popularity &0.1281&5.9110&0.1431&7.0332\\
			TextCNN &0.1127&6.9041&0.1301&8.9254\\
              ReDial &0.1185&6.8202&0.1242&8.9418\\
              KBRD &0.1149&5.8931&0.1222&9.0810\\
              KGSF &0.1055&6.7920&0.1198&9.1373\\
              KECRS&0.1072&6.8157&0.1075&9.0942\\
             \textbf{FacetCRS }&\textbf{0.0856} &\textbf{7.6436 }&\textbf{0.0902}&\textbf{10.8172 }\\
			\hline
		\end{tabular}
		\label{tab:filter_bubbles_}
	\end{center}
\end{table}

\begin{table}[h]
     \small
	\setlength{\tabcolsep}{6.5mm}
	\setlength{\abovecaptionskip}{0pt}  
	\begin{center}
		\renewcommand{\arraystretch}{1.0}
		\caption{Ablation studies on the conversation task in the REDIAL dataset.}
        \label{tab:each_component}
		\begin{tabular}{c|ccc}
             \hline
              Models&D-2&D-3&D-4\\
			\hline
			 \textbf{FacetCRS}&\textbf{0.126}&\textbf{0.209}&\textbf{0.305}\\
               w/o $\boldsymbol{X}_{\rm e}$ &0.118&0.201&0.298\\
               w/o $\boldsymbol{X}_{\rm w}$ &0.122&0.202&0.300\\
               w/o $\boldsymbol{X}_{\rm c}$ &0.119&0.197&0.296\\
               w/o $\boldsymbol{X}_{\rm r}$&0.123&0.205&0.301\\
			\hline
		\end{tabular}
		\label{tab:ablation}
	\end{center}
\end{table}

\subsection{Ablation Studies (RQ4)}
In this part, we conduct ablation experiments with different variants of our FacetCRS to verify the contributions of each component, including: 1) w/o $\boldsymbol{X}_{\rm e}$: we remove entity-facet embeddings $\boldsymbol{X}_{\rm e}$; 2) w/o $\boldsymbol{X}_{\rm w}$: we remove word-facet embeddings $\boldsymbol{X}_{\rm c}$; 3) w/o $\boldsymbol{X}_{\rm c}$: we remove context-facet embeddings $\boldsymbol{X}_{\rm c}$; 4) w/o $\boldsymbol{X}_{\rm r}$: we remove review-facet embeddings $\boldsymbol{X}_{\rm r}$. From Table \ref{tab:ablation}, we can see that removing any kind of various levels of preference facet leads to leads to performance degradation. This validates that these various data are beneficial for enhancing data representations to excavate diverse user preferences, thereby mitigating filter bubbles in the CRS to fulfill the recommendation goal.

\section{Conclusion and Future Work}
To prick filter bubbles in the CRS, we propose a novel framework FacetCRS, which models multi-faceted user preferences in the CRS, including entity-, word-, context-, and review-facet, to capture diverse user preferences to mitigate filter bubbles. Meanwhile, FacetCRS is an end-to-end framework to automatically learn representations of various levels of preference facet and diverse types of external knowledge. Through extensive experiments, our method consistently outperforms several competitive baselines, which demonstrate the effectiveness of our FacetCRS.

In the future, we will incorporate the causal reasoning into modeling different aspects of user preference. Additionally, we will also explore the causal relationships among various knowledge facets to further enhance CRS performance.

\section{Acknowledgments}
This work is supported in part by the National Key R\&D Program of China under Grant No. 2021ZD0111601, National Natural Science Foundation of China (NSFC) under Grant No. 61836012, 62325605, U21A20470, 62206110, 62206314, GuangDong Basic and Applied Basic Research Foundation under Grant No. 2023A1515011374, 2022A1515011835, China Postdoctoral Science Foundation funded project under Grant No. 2021M703687.

\bibliography{aaai24}
\end{document}